\documentclass[journal,10pt]{IEEEtran}
\IEEEoverridecommandlockouts
\usepackage{stfloats}
\usepackage{cite}

\usepackage{amsmath,amssymb,amsfonts}
\usepackage{threeparttable}
\usepackage{algorithm}
\usepackage{algorithmic}
\usepackage{graphicx}
\usepackage{textcomp}
\usepackage{xcolor}
\def\BibTeX{{\rm B\kern-.05em{\sc i\kern-.025em b}\kern-.08em      T\kern-.1667em\lower.7ex\hbox{E}\kern-.125emX}} 
\usepackage{verbatim}
\usepackage[utf8]{inputenc}
\usepackage{bbm}
\usepackage{array}
\usepackage{epstopdf}
\usepackage{multirow,url}
\usepackage{enumerate}
\usepackage{amsthm} 
\usepackage{bm}

\usepackage{booktabs} 
\usepackage{float} 

\title{A Persistence-Aware Framework for Age Violation Control in Wireless Status Update Systems
\author{Haoyuan~Pan,~\IEEEmembership{Member,~IEEE,} Chen~Chen, Shiyong~Zhou, Kun~Chen, and~Tse-Tin Chan,~\IEEEmembership{Member, IEEE}%
\thanks{H. Pan, C. Chen, S. Zhou, and K. Chen are with the College of Computer Science and Software Engineering, Shenzhen University, Shenzhen, China (e-mails:  {hypan@szu.edu.cn}; {2400101055@mails.szu.edu.cn}; {2510103056@mails.szu.edu.cn}; {2510105004@mails.szu.edu.cn}).}
\thanks{T.-T. Chan is with the Department of Mathematics and Information Technology, The Education University of Hong Kong, Hong Kong, China (e-mail: tsetinchan@eduhk.hk).}
}
}

\begin{document}

\maketitle
 
\begin{abstract}
Timely and reliable status updates are essential for emerging QoS-sensitive wireless applications. Common age of information (AoI)-based metrics, such as average AoI and age violation rate (AVR), characterize time-averaged freshness or violation frequency but do not explicitly capture the temporal persistence of consecutive age violations, which can be critical in safety-sensitive wireless applications. To address this limitation, we develop a persistence-aware reliability framework based on the \emph{consecutive age violation rate} (C-AVR) vector, whose components quantify AoI threshold violations over consecutive time windows of different lengths. Through flexible weighting schemes, the proposed framework unifies reliability objectives ranging from average persistence to tail-sensitive performance. Optimizing weighted C-AVR objectives is challenging because consecutive violations are temporally correlated, leading to sparse learning signals. To address this issue, we develop a distributional reinforcement learning approach based on a quantile regression dueling double deep Q-network (QR-D3QN). By modeling a quantile-based return distribution rather than only a scalar expected return, QR-D3QN provides richer value-estimation signals for rare but prolonged violation sequences under stochastic packet arrivals, unreliable channels, and transmission cost constraints. Simulation results demonstrate that QR-D3QN consistently outperforms expectation-based baselines across a wide range of weighting schemes and system settings, with particularly significant gains under tail-sensitive persistence objectives. Component-wise analysis further shows that distributional value learning substantially improves reliability across multiple persistence scales, especially for long consecutive violation sequences. Overall, our results establish the proposed C-AVR framework as an effective foundation for persistence-aware reliability evaluation and demonstrate the advantages of distributional reinforcement learning for wireless status update systems.
\end{abstract}

\begin{IEEEkeywords}
Age of information, distributional reinforcement learning, reliability, wireless scheduling.
\end{IEEEkeywords}

\section{Introduction}
Ensuring timely and reliable status updates is a fundamental requirement in wireless communication systems supporting industrial automation, autonomous driving, intelligent transportation, and real-time monitoring. In such applications, outdated information may degrade control accuracy, compromise safety, or destabilize closed-loop operation. To characterize information freshness, the age of information (AoI) has emerged as a widely adopted metric, measuring the time elapsed since the generation of the most recently received update \cite{AoI1,AoI2,AoI3}. Extensive studies have investigated AoI-aware scheduling, sampling, queue management, and resource allocation in wireless networks \cite{Op-AoI1,Op-AoI2,survey-AoI,pan2023timely}. Most existing works focus on average AoI or related mean-based freshness metrics due to their analytical tractability and intuitive interpretation of long-term system performance \cite{avg-AoI1,avg-AoI2,avg-survey2,DPP_3}.

However, average AoI is often insufficient for quality-of-service (QoS)-sensitive applications. In many practical systems, occasional but severe freshness degradation may dominate operational risk even when the average AoI remains acceptable. To address this limitation, reliability-oriented freshness metrics have been introduced. In particular, the age violation rate (AVR) measures the probability that AoI exceeds a predefined threshold and therefore provides a threshold-based characterization of timeliness reliability \cite{AVR1,AVR2,bounded_AoI}. Compared with average AoI, AVR is more directly connected to outage-like QoS guarantees and has been studied in wireless systems with stochastic arrivals, unreliable channels, and constrained scheduling resources \cite{AVR-dev1,AVR-dev2}. Nevertheless, existing violation-based metrics primarily quantify the frequency of violations, while largely ignoring their temporal persistence.

This limitation is particularly important in safety-critical systems. Consider two systems with the same AVR: in one system, violations occur sporadically and are rapidly corrected; in the other, violations occur in long uninterrupted runs. Although both systems exhibit identical violation probabilities, their operational risks can differ substantially. In applications such as autonomous control and industrial monitoring, prolonged information staleness may lead to sustained control errors, persistent sensing failures, or safety hazards \cite{safe1}. Therefore, reliability-aware freshness control should characterize not only whether AoI violations occur, but also how long they persist consecutively.

Motivated by this observation, we develop a persistence-aware reliability framework based on a vectorized characterization of consecutive AoI violations. Specifically, we introduce the \emph{consecutive age violation rate} (C-AVR) vector, whose $k$-th component measures the probability that AoI exceeds a threshold over $k$ consecutive time slots. This vector representation provides a \emph{persistence profile} of age violations across multiple time scales. Scalar reliability metrics arise as special cases under appropriate weightings on this vector; in particular, the scalar C-AVR for a fixed $k$ corresponds to a one-hot weighting. More generally, flexible weighting schemes over the C-AVR vector enable a unified treatment of reliability objectives, ranging from average persistence (uniform weighting) to tail-sensitive criteria that emphasize long violation runs (e.g., exponential weights). We refer to this scalar objective as the weighted C-AVR. \emph{To the best of our knowledge, this is the first AoI-based framework that models and optimizes consecutive violation persistence across multiple time scales.}

We study this framework in a multiuser wireless status update system, where a central scheduler selects users under transmission cost constraints, stochastic packet arrivals, and unreliable channels. Optimizing persistence-aware C-AVR objectives in this setting is challenging. Consecutive violation events are temporally coupled: once AoI exceeds the threshold, the violation tends to persist unless a successful update occurs. As a result, unlike average AoI or standard AVR, system performance depends not only on frequent and short violations, but also on rare and prolonged violation sequences. Moreover, system dynamics such as packet arrivals and channel reliability are often unknown or difficult to model accurately, limiting the applicability of classical model-based optimization methods \cite{model_sch_1,model_sch_2,model_sch_3}. 

These challenges motivate the use of deep reinforcement learning (DRL), which has recently emerged as an effective framework for wireless scheduling and resource allocation under uncertainty \cite{DRL+wireless-survey,AoI_RL}. Existing DRL-based wireless scheduling methods commonly employ expectation-based value-learning architectures, such as deep Q-networks (DQN) and their variants \cite{dou-DQN,dueling-DQN}. Such methods learn scheduling policies by optimizing the expected long-term return and have demonstrated strong performance in many communication and networking problems. DRL has also been applied to AoI-aware scheduling, particularly in scenarios where explicit system models are unavailable \cite{AoI_RL}. However, most existing DRL-based AoI studies focus on average AoI, average cost, or single-threshold reliability objectives, all of which can be adequately represented through scalar expected returns.

Persistence-aware reliability objectives fundamentally differ from these formulations. Weighted C-AVR objectives induce heterogeneous return distributions dominated by both frequent short violation runs and rare but prolonged persistence events. Under such objectives, expectation-based value learning may obscure critical tail behavior. Different scheduling actions may produce similar expected AoI performance while exhibiting substantially different risks of generating long consecutive violation sequences. Since expectation-based DRL represents the future return using only a scalar expectation, important persistence-related information in the return distribution may be obscured. This limitation becomes particularly pronounced under exponential weighting schemes, where long violation runs dominate system reliability.

To address this issue, we adopt distributional reinforcement learning, which models the full return distribution rather than only its expectation \cite{C51,QR-DQN,bdr2023}. By preserving information about return variability during value learning, distributional DRL is well-suited to persistence-aware reliability objectives whose rewards are induced by temporally correlated violation events. In particular, quantile-regression-based methods such as QR-DQN approximate the return distribution using learnable quantiles, enabling accurate modeling of asymmetric and high-variance return structures induced by consecutive violation events \cite{QR-DQN}. Recent studies have demonstrated the effectiveness of distributional DRL in uncertainty-aware wireless control, risk-sensitive scheduling, and reliability-constrained resource allocation \cite{dis-tail1,dis-tail2,dis-tail3,dis-tail4,disRLwork1,disRLwork2,disRLwork3,Constrained_RL}. Nevertheless, its role in persistence-aware AoI control has remained largely unexplored.

Building on these observations, we develop a quantile regression-based dueling double deep Q-network (QR-D3QN) tailored for persistence-aware wireless scheduling. The proposed architecture combines Double Q-learning to mitigate overestimation bias \cite{dou-DQN}, a dueling network structure to improve state-value and advantage estimation \cite{dueling-DQN}, and quantile regression to model the full return distribution \cite{QR-DQN}. These components jointly enable more stable learning under sparse and high-variance persistence-aware feedback, thereby reducing prolonged AoI violations while satisfying transmission cost constraints.

Extensive simulations suggest that \emph{multi-scale persistence objectives can benefit from distributional value representations, and that distributional DRL provides an effective learning approach}. Specifically, the proposed QR-D3QN consistently outperforms expectation-based baselines (i.e., DQN and D3QN) across a wide range of system configurations and weighting schemes, including uniform, exponential, and one-hot weighted C-AVR objectives. The performance gains become increasingly significant as the optimization objective places greater emphasis on long violation windows, confirming the importance of modeling the return distribution for persistence-aware reliability control. Furthermore, component-wise analysis of the C-AVR vector reveals that QR-D3QN achieves particularly substantial improvements at large persistence scales, demonstrating its effectiveness in suppressing prolonged consecutive AoI violations that dominate tail reliability. We further compare the proposed approach with the model-based drift-plus-penalty (DPP) method \cite{neely2010stochastic}. The results show that DRL methods achieve performance comparable to DPP under short persistence objectives, while QR-D3QN exhibits clear advantages when the objective emphasizes long violation windows. Moreover, all performance gains are achieved under the same transmission cost constraint, indicating that the improvements arise from more effective scheduling decisions rather than increased resource consumption. Overall, the results validate both the proposed persistence-aware C-AVR framework and the effectiveness of distributional DRL for reliability-aware wireless scheduling.

The main contributions of this work are summarized as follows:
\begin{itemize}
    \item We introduce the C-AVR vector to characterize AoI violations across multiple persistence scales, thereby capturing the full temporal persistence profile of information staleness. Based on this representation, we further develop the weighted C-AVR framework, which unifies average persistence, tail-sensitive reliability, and single-scale persistence objectives under flexible weighting schemes.
    
    \item We demonstrate empirically that persistence-aware C-AVR objectives induce heterogeneous and high-variance learning signals for which distributional value representations can be more effective than expectation-based DRL methods. To address this issue, we develop a distributional DRL framework based on QR-D3QN for persistence-aware wireless scheduling under stochastic arrivals, unreliable channels, and transmission cost constraints.
    
    \item We conduct extensive simulations under multiple weighting schemes and heterogeneous system settings. The results demonstrate that distributional value learning consistently outperforms expectation-based baselines, with particularly significant gains for long persistence windows and tail-sensitive reliability objectives. Component-wise analysis further confirms the effectiveness of the proposed method in suppressing prolonged AoI violation sequences across multiple temporal scales.
\end{itemize}

\section{System Model and Problem Formulation}\label{Sec-SMPF}
\subsection{System Model}\label{Sys-Mod}
\begin{figure}[t]
\centerline{\includegraphics[width=0.45\textwidth]{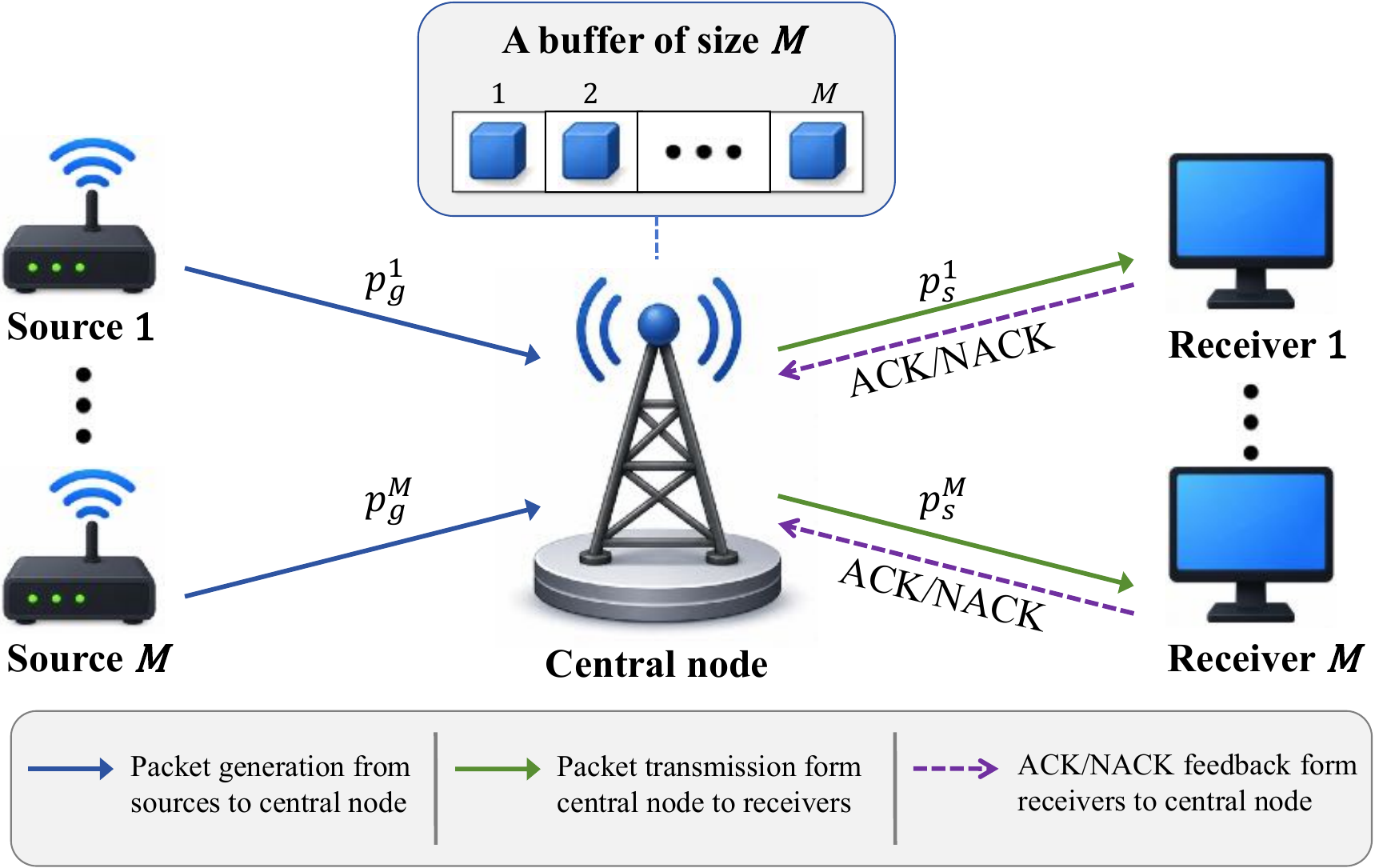}}
\caption{A time-slotted wireless status update system in which a central node schedules update packets to different receivers.}
\label{fig_System-Model}
\end{figure}

We consider a time-slotted wireless status update system with stringent information timeliness QoS requirements, consisting of $M$ independent information sources, indexed by $\{1,2,...,M\}$,  and their corresponding receivers coordinated by a central scheduler, as illustrated in Fig.~\ref{fig_System-Model}.

Time is indexed by discrete time slots $t=1,2,\dots$. At the beginning of each slot, every source independently generates at most one fresh status update packet according to a Bernoulli process, i.e., source $m$ generates a new packet with probability $p_g^m$. Newly generated packets are instantaneously delivered to the central node through an ideal uplink and overwrite any previously stored packet from the same source. Hence, the central node maintains a buffer of size $M$, consisting of $M$ per-source entries, with one entry per source storing only the most recent packet. This per-source replacement structure reflects QoS-driven designs where only the freshest information is relevant \cite{AoI3}.

Due to shared downlink resources, at most one packet can be transmitted in each time slot. A scheduled transmission to receiver $m$ succeeds with probability $p_s^m$, independently across slots.\footnote{For tractability, this work adopts a fixed transmission success probability model, which is commonly used in the AoI literature. This abstraction enables clear evaluation of scheduling and decision-making mechanisms without introducing additional physical-layer complexities. The proposed CMDP and DRL framework can be readily extended to more realistic time-varying wireless channels, such as Rayleigh fading or Markov fading channels, due to its model-free nature.} Upon a successful transmission, the receiver sends an acknowledgment (ACK) through an error-free feedback link; otherwise, a negative acknowledgment (NACK) is returned. 

The freshness of information is quantified using the age of information (AoI), defined as the time elapsed since the generation of the most recently successfully received status update. To track information timeliness across the system, AoI is maintained at both the transmitter and the receiver for each source $m \in \{1,2,...,M\}$:
\begin{itemize}
    \item At the central node, the transmitter-side AoI for source $m$ at slot $t$ is the number of slots since the most recent packet from source $m$ arrived at the buffer. The transmitter-side AoI for source $m$ at slot $t$, denoted by $\Delta_s(t,m)$, is $\Delta_s(t,m) = t - T_s(t,m)$, where $T_s(t,m)$ denotes the arrival time of the most recent packet from source $m$. The transmitter-side AoI is reset to zero upon a new packet arrival and increases by one in each subsequent slot otherwise.  

\item At the receiver, the receiver-side AoI $\Delta_r(t,m)$ measures the time elapsed since the generation of the most recently successfully received packet from source $m$. If an update packet from source $m$ is successfully delivered at the end of slot $t$, then $\Delta_r(t+1,m) = \Delta_s(t,m) + 1$; otherwise, $\Delta_r(t+1,m) = \Delta_r(t,m) + 1$. 
\end{itemize}

\begin{figure}[t]
\centerline{\includegraphics[width=0.45\textwidth]{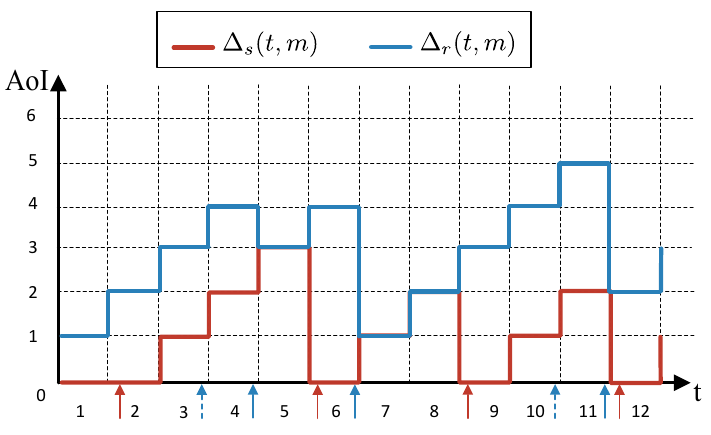}}
\caption{Sample AoI evolution for source $m$. The red curve shows the transmitter-side AoI $\Delta_s(t,m)$, and the blue curve shows the receiver-side AoI $\Delta_r(t,m)$. Red arrows indicate packet generation events at the beginning of slots, while blue arrows denote packet transmission outcomes at the end of slots: solid for successful transmissions and dashed for failures. }
\label{fig_AoI-Evolution}
\end{figure}

The transmitter-side and receiver-side AoI are initialized to zero and one, respectively. Their coupled evolution is illustrated via an example in Fig.~\ref{fig_AoI-Evolution}.  The transmitter-side AoI $\Delta_s(t,m)$ and the receiver-side AoI $\Delta_r(t,m)$ are depicted by the red curve and the blue curve, respectively. Red arrows indicate packet-generation events at the beginning of slots. Blue arrows represent packet transmission outcomes at the end of slots, where solid arrows correspond to successful transmissions and dashed arrows denote failures. 

As shown in Fig.~\ref{fig_AoI-Evolution}, the transmitter-side AoI is determined solely by packet generation events. Whenever a new packet is generated, $\Delta_s(t,m)$ is reset to zero, and otherwise it increases linearly by one in each subsequent slot. In contrast, the receiver-side AoI depends jointly on the transmitter-side AoI and the transmission outcome. When a packet transmission is successful at slot $t$, the receiver-side AoI at the next slot is $\Delta_r(t+1,m) = \Delta_s(t,m) + 1$, reflecting that the receiver obtains the most recently generated packet with a one slot delay. When a transmission attempt fails or when source $m$ is not scheduled, the receiver-side AoI simply increases by one. Fig.~\ref{fig_AoI-Evolution} demonstrates that consecutive transmission failures or scheduling delays can lead to sustained growth of receiver-side AoI even when fresh packets are available, highlighting the importance of capturing temporal persistence in reliability metrics.

\subsection{The Consecutive Age Violation Rate (C-AVR) Metric}\label{Pro-c-avr}
Most existing AoI-aware scheduling studies focus on long-term average performance metrics, with system average AoI being the most widely adopted measure of information freshness \cite{avg-survey2}. While effective for characterizing time-averaged performance, average AoI does not capture system reliability during rare but severe freshness-degradation events. To better capture reliability, recent works have introduced the age violation rate (AVR) metric, which measures the fraction of time slots in which the AoI exceeds a predefined threshold $\zeta$,
\begin{equation}
{\Phi}_r =
\limsup_{T \to \infty}
\frac{1}{MT}
\sum_{t=1}^{T}
\sum_{m=1}^{M}
\mathbb{E}
\left[
\mathbbm{1}{[ \Delta_r(t,m)>\zeta]}
\right].
\label{equ_AVR}
\end{equation}
where $\mathbbm{1}{[\cdot]}$ denotes the indicator function that returns 1 when its argument is true and 0 otherwise. Unlike average AoI, AVR captures the frequency of threshold violations rather than their magnitude, thereby providing a threshold-based notion of timeliness reliability.

However, AVR fundamentally treats violations as independent events and ignores their temporal structure. In particular, it does not distinguish between isolated threshold crossings and sustained violation periods, despite their markedly different implications for system performance. From a QoS and safety perspective, prolonged sequences of outdated information are significantly more harmful than sporadic and short-lived violations, as they may induce sustained decision errors or trigger system-level failures. This observation highlights a key limitation of AVR and motivates the need for a reliability metric that explicitly captures the temporal persistence of AoI violations.

To address this limitation, we put forth the \emph{consecutive age violation rate} (C-AVR), which quantifies the probability that the AoI exceeds the threshold across multiple consecutive time slots. Let a positive integer $k \ge 1$ denote a prescribed violation window length. A window of length $k$ ending at slot $t$ is said to be fully AoI violated for source $m$ if $\mathbbm{1}[\mathcal{V}_t^k(m)]=1$, where
\begin{equation}
\mathcal{V}_t^k(m)=\bigcap_{i=1}^{k} \{\Delta_r(t-k+i,m) > \zeta\}.
\end{equation}
That is, $\Delta_r(t-k+i,m) > \zeta$ for all $i=1,\cdots,k$. 

From a QoS standpoint, C-AVR characterizes the probability of timeliness outage events whose duration exceeds $k$ consecutive slots, thereby directly capturing persistence in reliability degradation. The system-level C-AVR is defined as
\begin{equation}
{\Psi}_r^{k}=
\limsup_{T \to \infty}
\frac{1}{M(T-k+1)}
\sum_{t=k}^{T}
\sum_{m=1}^{M}
\mathbb{E}
\left[
\mathbbm{1}[\mathcal{V}_t^k(m)]
\right].
\label{equ_C-AVR}
\end{equation}
Optimizing C-AVR therefore enforces temporal QoS guarantees on information freshness, rather than merely improving average performance. C-AVR naturally recovers AVR as a special case when $k=1$, while larger values of $k$ place increasing emphasis on severe reliability degradation caused by prolonged violation sequences. Furthermore, since C-AVR depends on the joint temporal distribution of AoI exceedances, it cannot be inferred from AVR alone. This highlights that persistence is an intrinsic dimension of reliability that is not captured by existing metrics. To the best of our knowledge, no such window-based formulation that explicitly characterizes violation persistence has been systematically developed in the AoI scheduling literature.

\subsection{C-AVR Vector and Weighted Reliability Objective}\label{Pro-vector-c-avr}
While the C-AVR defined in \eqref{equ_C-AVR} captures persistence over a fixed window length $k$, it remains inherently a single-scale characterization. In practice, however, violations of different durations correspond to different levels of operational risk. Short consecutive violations may be tolerable in certain applications, whereas long violation runs can lead to sustained performance degradation or even system instability. Therefore, a metric defined at a single persistence scale is insufficient to fully characterize system reliability.

We introduce a vectorized characterization of persistence, called the \emph{C-AVR vector}, in which each component corresponds to a distinct violation window length. Specifically, for a maximum window length $k_{\max}$, the C-AVR vector is defined as
\begin{equation}
\mathbf{\Psi} = \left(\Psi_r^1,\Psi_r^2,\ldots,\Psi_r^{k_{\max}}\right),
\label{equ_vector_cavr}
\end{equation}
This representation provides a multi-scale persistence profile of AoI violations, capturing reliability behavior across a spectrum of temporal horizons rather than at a single scale. As such, it elevates C-AVR from an isolated scalar metric to a structured description of temporal reliability.

To enable tractable optimization while preserving this multi-scale structure, we introduce a weighted aggregation of the C-AVR vector. Let $\mathbf{W}=(w_1,w_2,...,w_{k_{\max}})$ denote the weighting vector satisfying $w_k \geq 0$ and $\sum_{k=1}^{k_{\max}} w_k = 1$, where $w_k$ specifies the relative importance assigned to persistence at scale $k$, i.e., the C-AVR scalar ${\Psi}_r^{k}$. The resulting scalar objective, referred to as the \emph{weighted C-AVR}, is defined as
\begin{equation}
\bar{\Psi} = \mathbf{W} \mathbf{\Psi}^\top=\sum_{k=1}^{k_{\max}} w_k \Psi_r^k,
\label{equ_weighted_cavr}
\end{equation}

Importantly, this formulation establishes a unified optimization framework over the persistence profile. By adjusting the weights, the scheduler can emphasize different persistence regimes, ranging from average persistence to tail-sensitive criteria that prioritize long violation runs. In this sense, conventional scalar formulations arise as special cases within the proposed framework. In particular, a one-hot weighting reduces the weighted C-AVR to a single-scale C-AVR metric, while more general weightings capture interactions across multiple persistence scales. In this work, we consider the following representative weighting schemes:
\begin{itemize}
\item \textbf{Uniform Weights} assign equal importance across all persistence scales, i.e.,
\begin{equation}
w_k = \frac{1}{k_{\max}}, \quad k=1,2,\ldots,k_{\max},
\end{equation}
thus reflecting scenarios where both short and long violation sequences are considered equally relevant.

\item \textbf{Exponential Weights} assign progressively larger weights to longer violation windows, i.e.,
\begin{equation}
w_k = \frac{\beta^k}{\sum_{j=1}^{k_{\max}} \beta^j}, \quad \beta > 1,
\end{equation}
thereby emphasizing tail behavior. Consequently, the resulting weighted objective is particularly sensitive to long but rare violations, with $\beta$ controlling the degree of tail emphasis.

\item \textbf{One-hot Weights} concentrate all weight on a single persistence scale $k_o$, i.e., $w_{k_o}=1$ and $w_k=0$ for $k \neq k_o$. In this case, the weighted C-AVR reduces exactly to the scalar C-AVR at window length $k_o$, demonstrating that the scalar formulation is a special case of the proposed vector framework.
\end{itemize}

\subsection{Problem Formulation}\label{Pro-For}
Let $a_t \in \{0,1,...,M\}$ denote the scheduling action at slot $t$, where $a_t = m$ indicates that source $m$ is scheduled for transmission and $a_t = 0$ indicates no transmission. The long-term average transmission cost is defined as
\begin{equation}
\eta =
\limsup_{T\to\infty}
\frac{1}{T}
\sum_{t=1}^{T}
\mathbb{E}[\mathbbm{1}[a_t\neq 0]].
\end{equation}
which must satisfy $\eta \leq \eta_{\max}$, where $\eta_{\max} \in (0,1]$ reflects the resource constraint.

Combining the persistence-aware reliability objective with the resource constraint, we formulate the weighted C-AVR scheduling problem as
\begin{equation}
\begin{aligned}
& \min_{\{a_t\}} ~~\bar{\Psi}, \\
& \ \text{s.t.} \quad \eta \leq \eta_{\max}. 
\end{aligned}\label{Problem}
\end{equation}

By optimizing the weighted C-AVR, the scheduler is explicitly driven to control the temporal persistence of AoI violations, rather than merely their instantaneous occurrence. This distinction is critical because unlike AVR, which primarily reflects frequency-based characteristics, the weighted C-AVR objective is inherently sensitive to the temporal correlation structure of violations and is therefore dominated by rare but high-impact events.
Furthermore, the weighting mechanism enables a unified treatment of reliability objectives across multiple persistence scales. In particular, uniform weighting captures average persistence behavior, exponential weighting emphasizes tail risk, and one-hot weighting isolates single-scale reliability requirements. This unified formulation highlights the conceptual contribution of the proposed framework: reliability is not characterized by a single metric, but by a structured persistence profile that can be flexibly aggregated based on system requirements. In the next section, we cast problem \eqref{Problem} as a constrained Markov decision process (CMDP), which serves as the foundation for the proposed distributional reinforcement learning approach.

\section{CMDP Formulation}\label{Sec-CMDP}
The scheduling problem formulated in Section~\ref{Pro-For} involves long-term persistence-aware reliability optimization under an average transmission cost constraint. Such a problem can be naturally modeled as a CMDP \cite{cmdp}. In this section, we present a complete CMDP formulation that is explicitly aligned with the weighted C-AVR objective introduced previously. The CMDP is defined as a sextuple $(\mathcal{S}, \mathcal{A}, \mathcal{P}, \mathcal{R}, \mathcal{C}, \gamma)$, where $\mathcal{S}$ is the state space, $\mathcal{A}$ is the action space, $\mathcal{P}: \mathcal{S} \times \mathcal{A} \times \mathcal{S} \rightarrow [0,1]$ is the state transition probability, $\mathcal{R}: \mathcal{S}\times\mathcal{A} \rightarrow \mathbb{R}$ is the reward function, $\mathcal{C}: \mathcal{S}\times\mathcal{A} \rightarrow \mathbb{R}$ is the cost function, and $\gamma$ is the discount factor. Each component is specified in detail below.

\paragraph{State Space}
The system state captures both the instantaneous information freshness and its temporal persistence at a given time slot. Since both packet generation and packet delivery affect AoI evolution, the state jointly includes the transmitter-side AoI, the receiver-side AoI, and the consecutive violation status for each source $m$. At the beginning of slot $t$, the system state is represented by a $3M$-dimensional vector
\begin{align}
s_t = \big(&\Delta_s(t,1), \Delta_r(t,1), v(t,1), \ldots, \notag \\
&\Delta_s(t,m), \Delta_r(t,m), v(t,m), \ldots, \notag \\
&\Delta_s(t,M), \Delta_r(t,M), v(t,M)\big).
\end{align}
Here, $\Delta_s(t,m)$ and $\Delta_r(t,m)$ denote the transmitter-side and receiver-side AoI of source $m$, respectively. The variable $v(t,m)$ tracks the length of the current consecutive AoI violation segment relative to the threshold $\zeta$, evolving as
\begin{equation}
v(t,m) =
\begin{cases}
v(t-1,m) + 1, & \text{if } \Delta_r(t,m) > \zeta, \\
0, & \text{if } \Delta_r(t,m) \le \zeta,
\end{cases}
\end{equation}
with initial condition $v(0,m) = 0$. $v(t,m)$ is reset whenever the AoI returns to a non-violating level. To ensure a finite-state MDP, the violation counter is truncated at $k_{\max}$ in this work, i.e., $v(t,m)\in\{0,1,\dots,k_{\max}\}$. Specifically, whenever the number of consecutive AoI violations exceeds $k_{\max}$, the counter remains fixed at $k_{\max}$.

This state representation explicitly encodes both instantaneous AoI levels and their temporal persistence, which is essential for evaluating multi-scale violation events underlying the C-AVR vector. To ensure a finite state space, all AoI values are truncated by a predefined upper bound $\Delta_{\max}$. Once an AoI value reaches $\Delta_{\max}$, it remains at this value. This truncation preserves the relevant persistence structure while improving computational tractability.

\paragraph{Action Space}
At each time slot $t$, the scheduler selects an action $a_t \in \mathcal{A}$, where $\mathcal{A} = \{0, 1, \ldots, M\}$. Action $a_t = m$ schedules source $m$ for transmission, while $a_t=0$ indicates no transmission in the current slot.

\paragraph{State Transition Probability}
Given the current state $s_t$ and action $a_t$, the system moves to the next state $s_{t+1}$ according to the transition probability $\mathcal{P}(s_{t+1}\mid s_t,a_t)$. The transition is determined by stochastic packet generation and random transmission outcomes. Specifically, packet generation for source $m$ follows a Bernoulli process with probability $p_g^m$, and successful transmission occurs with probability $p_s^m$. These system parameters are assumed unknown, making the transition model unavailable in closed form. Consequently, the scheduling problem naturally motivates a model-free reinforcement learning approach.

\paragraph{Cost Function}
To model resource constraints, each transmission incurs a unit cost that captures resource consumption, such as energy or bandwidth. The instantaneous cost at slot $t$ is defined as $c_t=\mathbbm{1}[a_t \neq 0]$. The long-term average transmission cost is required to satisfy $\lim_{T \to \infty} \frac{1}{T} \sum_{t=1}^{T} c_t \le \eta_{\max}$, where $\eta_{\max} \in (0,1]$ specifies the allowable resource budget.

\paragraph{Reward Function}
The objective is to minimize the weighted C-AVR defined in \eqref{equ_weighted_cavr}, which aggregates violation events across multiple persistence scales. To incorporate this objective into the CMDP framework while satisfying the resource constraint, we adopt a Lagrangian relaxation approach. Specifically, after executing action $a_t$ during slot $t$, the reward is defined as
\begin{equation}
\begin{aligned}
r_t\triangleq
-\frac{1}{M}\sum_{k=1}^{k_{\max}} w_k \sum_{m=1}^{M}
\mathbbm{1}[\mathcal{V}_{t+1}^k(m)]
-\lambda(c_t-\eta_{\max}).
\end{aligned}
\label{equ_reward}
\end{equation}
Here, $\mathcal{V}_{t+1}^k(m)$ denotes the event that the receiver-side AoI of source $m$ exceeds the threshold $\zeta$ for $k$ consecutive slots ending at slot $t+1$. The index $t+1$ appears because the scheduling action taken in slot $t$ affects the AoI state only after the transmission outcome is realized at the end of the slot. Consequently, the persistence-aware violation event induced by action $a_t$ is evaluated using the updated AoI state at slot $t+1$. The second term in \eqref{equ_reward} enforces the transmission cost constraint through the Lagrange multiplier $\lambda \geq 0$. The first term represents a weighted aggregation of multi-scale violation events, and its time average corresponds to the empirical estimate of the weighted C-AVR defined in \eqref{equ_weighted_cavr}. It therefore penalizes violations across multiple persistence scales simultaneously, with their relative importance determined by the weight vector $\mathbf{W}$. For example, uniform weights emphasize average persistence behavior, whereas exponentially increasing weights place greater emphasis on rare but prolonged violation runs associated with tail reliability degradation.

The empirical average transmission cost is updated as
\begin{equation}
\eta \leftarrow \eta + \frac{1}{t}\big(c_t-\eta\big),
\label{equ_eta_t}
\end{equation}
and the Lagrange multiplier $\lambda$ is updated using the projected stochastic subgradient descent
\begin{equation}
\lambda \leftarrow \max\left(0,\lambda+\xi(\eta-\eta_{\max})\right),
\label{equ_lambda}
\end{equation}
where $\xi>0$ is the step size. This update dynamically adjusts the penalty to enforce the long-term constraint, i.e., increasing the penalty when the empirical transmission cost exceeds the allowable budget.

\paragraph{{Discount Factor}}
The discount factor $\gamma \in (0,1]$ controls the relative importance of immediate and future rewards. When the system performance is evaluated through the expected discounted return
$\mathbb{E}_\pi\left[\sum_{t=0}^{\infty} \gamma^t r_t \right]$ under policy $\pi$, a value of $\gamma$ close to one emphasizes long-term reliability performance, which is critical for persistence-aware objectives where the impact of actions accumulates over consecutive time slots.

While the CMDP formulation provides a principled framework for optimizing the weighted C-AVR objective under resource constraints, it also exposes a key challenge. The weighted C-AVR objective is affected by temporally correlated and potentially rare violation events, especially under tail-sensitive weighting schemes. Consequently, the induced return distribution can be high-variance and asymmetric. When future performance is summarized solely by its expectation, some persistence-driven variability is compressed into a scalar value estimate. In particular, conventional DRL methods based on expectation-valued value functions may yield less informative learning signals for prolonged violation sequences under persistence-aware objectives. This observation motivates value representations that preserve distributional information during training. In the next section, we develop a distributional DRL framework tailored to weighted C-AVR optimization.

\section{Distributional DRL for Weighted C-AVR Optimization}\label{Sec-dist}
\begin{figure*}[t]
\centerline{\includegraphics[width=0.93\textwidth]{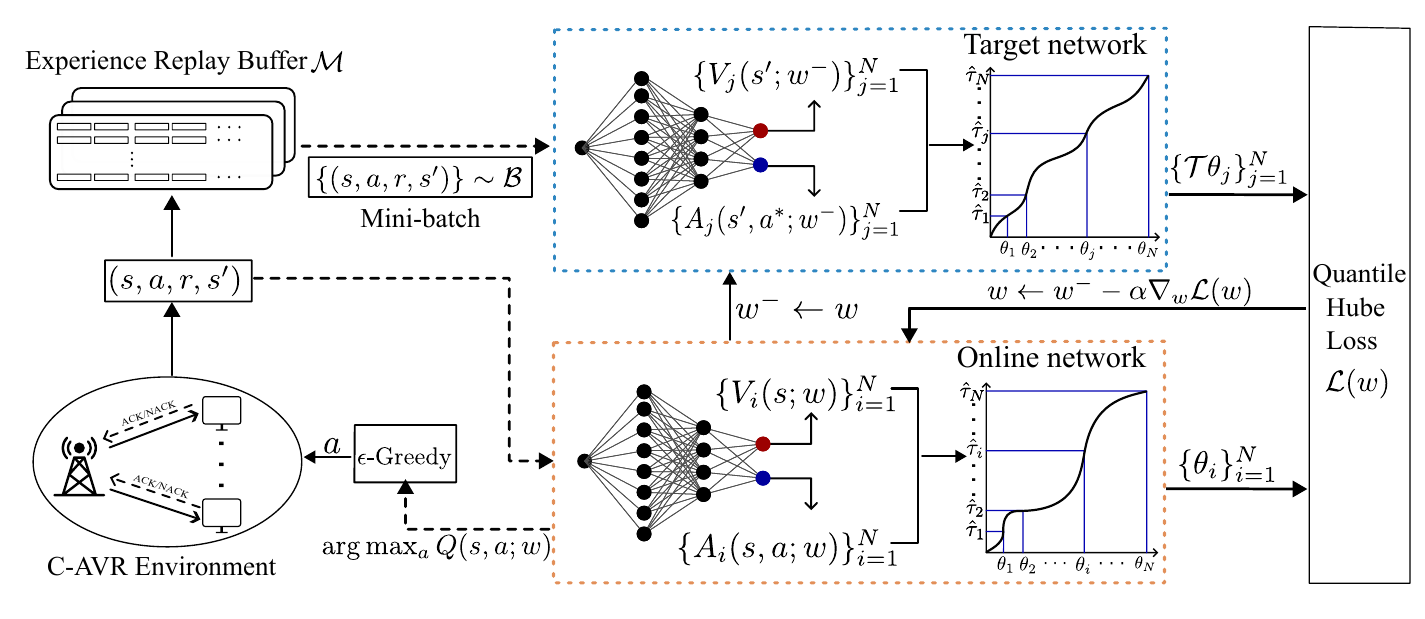}}
\caption{QR-D3QN–based training framework for weighted C-AVR–aware scheduling, where distributional value estimation enables learning under rare but persistent AoI violation events.}
\label{fig-Framework}
\end{figure*}

This section develops a distributional DRL approach explicitly aligned with the weighted C-AVR objective. By modeling the full distribution of future returns, rather than only their expectation, the proposed method captures the return variability and tail behavior induced by persistence-aware AoI violation penalties across multiple time scales. This richer representation enables the scheduler to distinguish policies with similar average performance but markedly different reliability profiles. Leveraging quantile-based distributional learning together with a reliability-aware state and reward design introduced in the CMDP formulation, we construct a QR-D3QN framework that is naturally suited to minimizing persistence-driven QoS violations under resource constraints. The overall learning framework is illustrated in Fig.~\ref{fig-Framework}. 

\subsection{From Expected Value Learning to Distributional Value Learning}
In standard Deep Q-Networks (DQN), the agent learns a scalar action-value function $Q(s,a)$, representing the expected discounted return obtained by taking action $a$ in state $s$. The optimal action-value function satisfies the Bellman optimality equation
\begin{equation}
Q(s,a)
= \mathbb{E}\left[ r + \gamma \max_{a'} Q(s',a') \mid s,a \right],
\end{equation}
where $r$ is the immediate reward, $s'$ is the next state, and $\gamma \in (0,1]$ is the discount factor. 

In practice, $Q(s,a)$ is approximated by a neural network $Q(s,a;w)$ with parameters $w$, trained by minimizing the temporal-difference (TD) loss 
\begin{equation}
\mathcal{L}(w) = \mathbb{E}_{(s,a,r,s')\sim \mathcal{B}} \Big[ \big( y - Q(s,a;w) \big)^2 \Big],
\end{equation}
where the TD target is
\begin{equation}
y = r + \gamma \max_{a'} Q(s',a'; w^-), 
\end{equation}
and $w^-$ denotes the parameters of a periodically updated target network. Here, $ \mathcal{B}$ represents mini-batches sampled uniformly from the replay buffer $\mathcal{M}$, which mitigates temporal correlations and stabilizes learning. 

A known limitation of DQN is the overestimation bias introduced by the maximization operator. Dueling Double DQN (D3QN) mitigates this issue through two complementary mechanisms:
\begin{itemize}
    \item \textbf{Double Q-learning}: Action selection and evaluation are decoupled, yielding the TD target
\begin{equation}
y^* = r + \gamma Q\Big(s', \arg\max_{a'} Q(s',a'; w) ; w^-\Big).
\end{equation}
    \item \textbf{Dueling architecture}: The Q-function is decomposed into a state-value component and an advantage component
\begin{equation}
Q(s,a; w) = V(s;w) + \Big( A(s,a;w) - \frac{1}{|\mathcal{A}|} \sum_{a'} A(s,a';w) \Big),
\end{equation}
where $|\mathcal{A}|$ is the number of actions. This separation improves stability and emphasizes meaningful actions relative to the state.
\end{itemize}

While D3QN improves learning stability and action discrimination, it still estimates only the expected return. This limitation is particularly relevant for weighted C-AVR optimization. Specifically, the reward in \eqref{equ_reward} aggregates multi-scale violation events and is influenced by temporally correlated and potentially rare trajectories involving persistent AoI violations, especially under tail-sensitive weighting schemes. As a result, the induced return distribution can be heterogeneous and high-variance. When learning is based solely on expected returns, these persistence-driven variations are compressed into scalar value targets. In particular, actions or policies with similar expected returns may still be associated with return distributions of different variability, which can make expectation-based value learning less informative under persistence-aware reliability objectives.

To address this challenge, we adopt a distributional reinforcement learning framework. Instead of approximating the expected return, distributional methods model the entire return distribution. Let $Z(s,a)$ denote the random return obtained by taking action $a$ in state $s$. The expected action value is then
\begin{equation}
Q(s,a) = \mathbb{E}[Z(s,a)].
\end{equation}
By explicitly representing the distribution of returns $Z(s,a)$, distributional DRL preserves information about variability and low-probability, high-impact events, which are precisely the events governing weighted C-AVR performance.

\subsection{QR-D3QN for C-AVR-based Scheduling}
Among distributional DRL approaches, Quantile Regression DQN (QR-DQN) offers a favorable balance between modeling flexibility and computational efficiency. Rather than using fixed categorical supports in C51, QR-DQN approximates the return distribution using a set of learnable quantiles.

For each state action pair $(s,a)$, the cumulative distribution function (CDF) is partitioned into $N$ uniform probability intervals $\{\tau_1, \tau_2,..., \tau_i,..., \tau_N\}$, where $\tau_i=i/N$. The midpoint of the interval $(\tau_{i-1}, \tau_i)$ is $\hat{\tau}_i = (\tau_{i-1}+\tau_i)/2$. The network learns quantile values $\theta_i(s,a; w)$ representing the $\hat{\tau}_i$-quantiles, and the return distribution of $Z(s,a; w)$ is approximated by a uniform mixture of Dirac masses at these learned quantiles
\begin{equation}
Z(s,a; w)
= \frac{1}{N} \sum_{i=1}^{N}
\delta_{\theta_i(s,a; w)},
\end{equation}
where $\delta_z$ denotes a Dirac distribution centered at $z$. This representation captures not only the central tendency of the return but also its dispersion and tail behavior. In the context of weighted C-AVR optimization, the reward explicitly penalizes persistence of AoI violations across multiple time scales. Consequently, trajectories containing prolonged violation sequences accumulate substantially larger penalties and therefore induce significantly lower cumulative returns. These rare but high-impact trajectories primarily influence the lower quantiles of the learned return distribution. Consequently, the distributional representation preserves richer training information about persistence-driven reliability degradation, whereas expectation-based methods compress such information into a single scalar target.

To further improve learning stability, we integrate quantile regression with the D3QN architecture. For each quantile index $i \in \{1,\ldots,N\}$, the corresponding quantile value $\theta_i(s,a;w)$ is decomposed into a state-value component and an advantage component as
\begin{equation}
\theta_i(s,a;w)
= V_i(s;w)
+ \Big(
A_i(s,a;w)
- \frac{1}{|\mathcal{A}|} \sum_{a'} A_i(s,a';w)
\Big),
\label{equ_theta}
\end{equation}
with
\begin{align}
V_i(s;w) &= \frac{1}{|\mathcal{A}|} \sum_{a'} \theta_i(s,a';w), \\
A_i(s,a;w) &= \theta_i(s,a;w) - V_i(s;w). 
\end{align}
This decomposition enforces $\sum_{a} A_i(s,a;w)=0$ for each quantile $i$, yielding a consistent separation between state-value and action-dependent effects at every quantile level. It extends the classical dueling architecture to the distributional setting, where the components ${V_i(s)}$ capture the system-wide reliability baseline at different return quantiles, while ${A_i(s,a)}$ quantify the action-specific deviations relative to this baseline. The quantile-wise dueling architecture thus enables the model to distinguish how different actions influence not only the expected performance but also the distributional characteristics of returns, including variability and tail risk induced by persistent violation events.

For action selection, the expected action value is computed by averaging the quantile estimates 
\begin{equation}
Q(s,a;w) = \frac{1}{N} \sum_{i=1}^{N} \theta_i(s,a;w),
\end{equation}

In QR-D3QN, the mean action value used for scheduling is obtained by averaging the learned quantile returns. The key distinction from expectation-based DRL lies in how this action value is learned. Instead of directly fitting a single scalar TD target, QR-D3QN learns a set of quantile targets through distributional Bellman updates, thereby preserving information about return variability, asymmetry, and low-probability persistence events during training. As a result, the resulting mean action-value estimates are informed by a richer distributional representation, which helps the scheduler identify actions that reduce prolonged AoI violation sequences under high-variance persistence-aware rewards. 

In addition, actions are selected according to an $\epsilon$-greedy exploration policy, where the exploration rate $\epsilon_e$ at episode $e$ follows the linear decay schedule
\begin{equation}
\epsilon_e =
\begin{cases}
\epsilon_{\mathrm{start}} -
\dfrac{e-1}{E_{\mathrm{decay}}-1}
(\epsilon_{\mathrm{start}}-\epsilon_{\mathrm{end}}),
& e \leq E_{\mathrm{decay}}, \\
\epsilon_{\mathrm{end}}, & e > E_{\mathrm{decay}}.
\end{cases}
\label{eqn:exploration_rates}
\end{equation}
Here, $\epsilon_{\mathrm{start}}$ and $\epsilon_{\mathrm{end}}$ denote the initial and final exploration rates, respectively. $E_{\mathrm{decay}}$ specifies the number of training episodes over which the exploration rate decreases linearly from $\epsilon_{\mathrm{start}}$ to $\epsilon_{\mathrm{end}}$.

\begin{algorithm}[t]
\caption{QR-D3QN for Weighted C-AVR Scheduling}
\label{alg:QR-D3QN-C-AVR}
\begin{algorithmic}[1]
\REQUIRE 
Number of training episodes $E_{\max}$; 
episode length $T_{\max}$;
minimum replay buffer size $\mathcal{M}_{\min}$ for training;
discount factor $\gamma$;  
learning rate $\alpha$;  
target update frequency $G$; 
linear decay exploration schedule parameters $(\epsilon_{\mathrm{start}}, \epsilon_{\mathrm{end}}, E_{\mathrm{decay}})$;
transmission cost constraint $\eta_{\max}$;  
Lagrange multiplier $\lambda$;  
number of quantiles $N$;  
step size $\xi$ for $\lambda$ update.


\STATE Initialize online network $w$ and target network $w^- \leftarrow w$, replay buffer $\mathcal{M}$, Lagrange multiplier $\lambda \ge 0$, empirical cost estimate $\eta \gets 0$, and global time slot $t\gets 0$

\FOR{episode $e = 1$ to $E_{\max}$}
    \STATE Reset the environment
    \STATE Update exploration rate $\epsilon_e$ according to the linear decay schedule \eqref{eqn:exploration_rates}
    \FOR{$n=1$ to $T_{\max}$}
        \STATE Update global time slot index $t\gets t+1$
        \STATE Observe the current state $s_{t}$ consisting of AoI states and consecutive violation counters
        \STATE Compute quantile values $\{\theta_i(s_{t},a;w)\}_{i=1}^{N}$ for all $a \in \mathcal{A}$ using the quantile-wise dueling architecture (\ref{equ_theta})
        \STATE Compute mean action values $Q(s_{t},a;w) = \frac{1}{N} \sum_{i=1}^{N} \theta_i(s_{t},a;w)$
        \STATE Select action $a_{t}$ using $\epsilon_e$-greedy exploration
        \STATE Execute $a_{t}$, observe cost $c_{t}$, compute the Lagrangian reward $r_{t}$ using the current multiplier $\lambda$, and obtain the next state $s_{t+1}$
        \STATE Store transition $(s_{t},a_{t},r_{t},s_{t+1})$ in $\mathcal{M}$ 
        \STATE Update empirical average transmission cost estimate $\eta$ by \eqref{equ_eta_t}
        \IF{$|\mathcal{M}| \ge \mathcal{M}_{\min}$}
            \STATE Sample a mini-batch of transitions  $\mathcal{B} = \{(s,a,r,s')\}$ uniformly from the replay buffer $\mathcal{M}$
            \STATE Select the greedy next action using the online network by (\ref{equ_action_sel})
            \STATE Compute target quantiles by (\ref{equ_target_quantile})
            \STATE Compute quantile Huber loss $\mathcal{L}(w)$ by (\ref{equ:huber_loss})
            \STATE Gradient update: $w \gets w - \alpha \nabla_w \mathcal{L}(w)$
            \IF{$t \bmod G = 0$}
                \STATE Update target network: $w^- \gets w$
            \ENDIF
        \ENDIF
        \STATE Update Lagrange multiplier $\lambda$ by (\ref{equ_lambda})
    \ENDFOR
\ENDFOR
\RETURN Trained QR-D3QN parameters $w$
\end{algorithmic}
\end{algorithm}

The distributional Bellman update follows the double Q-learning principle. For a sampled transition $(s,a,r,s')$, the next action is selected as
\begin{equation}
a^* = \arg\max_{a'} \frac{1}{N} \sum_{i=1}^{N} \theta_i(s',a'; w), 
\label{equ_action_sel}
\end{equation}
and the target quantiles are computed as
\begin{equation}
\mathcal{T}\theta_j = r + \gamma \theta_j(s',a^*; w^-), \quad j=1,\dots,N,
\label{equ_target_quantile}
\end{equation}
where $w$ and $w^-$ denote the online and target network parameters, respectively. 

To train the network parameters, we minimize the quantile Huber loss
\begin{equation}
\mathcal{L}(w) = \frac{1}{N} \sum_{i=1}^{N} \sum_{j=1}^{N} \rho_{\hat{\tau}_i}^\kappa\big( \mathcal{T}\theta_j - \theta_i(s,a; w) \big),
\label{equ:huber_loss}
\end{equation}
with
\begin{equation}
\rho_{\tau}^\kappa(u) =
\begin{cases}
\frac{1}{2} u^2 |\tau - \mathbbm{1}\{u < 0\}|, & |u| \le \kappa, \\
\kappa(|u| - \frac{1}{2}\kappa) |\tau - \mathbbm{1}\{u < 0\}|, & \text{otherwise}.
\end{cases}
\end{equation}
where $\kappa>0$ controls smoothness near zero. This objective ensures accurate learning of the return distribution while maintaining training stability.

As shown in Fig.~\ref{fig-Framework}, the proposed QR-D3QN framework integrates quantile-based distributional value learning with the D3QN learning architecture to optimize the weighted C-AVR objective in stochastic wireless systems. At each time slot, the agent observes the system state $s$, which includes transmitter-side AoI, receiver-side AoI, and consecutive violation counters, and selects an action $a$ via an $\epsilon$-greedy policy based on the expected Q-values computed by averaging the learned quantiles. After executing the action, the resulting transition $(s,a,r,s')$ is stored in the replay buffer. Mini-batches sampled from this buffer are used to update the online network by minimizing the quantile Huber loss, where target quantiles are constructed through a distributional Bellman update combined with double Q-learning. A target network is periodically synchronized to ensure training stability. 

Crucially, by modeling a quantile-based return distribution during training, the learned value representation retains information related to both the frequency of AoI violations and their temporal persistence across different scales. This can improve the mean action-value estimates used by the mean-greedy scheduler when returns are high-variance and persistence-aware. Such capability is useful for weighted C-AVR optimization, especially under weighting schemes that place larger penalties on prolonged violation events. The complete procedure is summarized in Algorithm~\ref{alg:QR-D3QN-C-AVR}.


\section{Performance Evaluation}\label{Sec-PE}
This section evaluates the proposed persistence-aware scheduling framework from three perspectives: (i) the effectiveness of the weighted C-AVR objective in capturing multi-scale reliability, (ii) the benefit of distributional reinforcement learning in reducing persistent AoI violations, and (iii) the robustness of the proposed method under varying system conditions.

\subsection{Simulation Setup}\label{Sec-PE1}
All algorithms are implemented in Python 3.9 using PyTorch and evaluated under identical software and hardware configurations to ensure a fair and reproducible comparison. Unless otherwise specified, we consider a system with $M=10$ sources, and a homogeneous setting for different sources with packet generation probability $p_g^m=0.7$, transmission success probability $p_s^m=0.7$, AoI violation threshold $\zeta=15$, and transmission cost constraint $\eta_{\max}=0.75$. Heterogeneous settings are considered in Section \ref{Sec-PE4}. The AoI is truncated at $\Delta_{\max}=100$. The training parameters are summarized in Table~\ref{tab_parameter}: discount factor $\gamma=0.98$, learning rate $\alpha=2\times10^{-3}$, replay buffer size $|\mathcal{M}|=10^6$, minimum buffer size $\mathcal{M}_{\min}=1000$, mini-batch size $|\mathcal{B}|=256$, target update period $G=3$, and number of quantiles $N=64$. All results are averaged over 10 independent runs with different random seeds.

For weighted C-AVR simulations, we set $k_{\max}=9$ unless otherwise stated. Three weighting schemes are considered: (i) Uniform weights; (ii) Exponentially increasing weights (default $\beta=2$); and (iii) One-hot weights, which reduce the objective to a single-window C-AVR.

\begin{table}[t]
\centering
\caption{Training Parameters of the QR-D3QN Algorithms}
\label{tab_parameter}
\begin{tabular}{lcc}
\hline
\textbf{Parameter} & \textbf{Symbol} & \textbf{Value} \\
\hline
AoI upper bound & $\Delta_{\max}$ & $100$ \\
Discount factor & $\gamma$ & $0.98$ \\
DRL learning rate & $\alpha$ & $2\times10^{-3}$ \\
Step size for $\lambda$ update & $\xi$ & $0.10$ \\
Initial exploration rate & $\epsilon_{\mathrm{start}}$ & $1.0$ \\
Final exploration rate & $\epsilon_{\mathrm{end}}$ & $0.05$ \\
Number of exploration decay episodes & $E_{\mathrm{decay}}$ & $1000$ \\
Target network update period & $G$ & $3$ \\
Replay buffer size & $|\mathcal{M}|$ & $10^{6}$ \\
Minimum replay buffer size for training & $\mathcal{M}_{\min}$ & $1000$ \\
Mini-batch size & $|\mathcal{B}|$ & $256$ \\
Hidden-layer size & -- & $2\times128$ \\
Number of quantiles & $N$ & $64$ \\
Huber loss threshold & $\kappa$ & $1$ \\
Number of training episodes & $E_{\max}$ & $3000$ \\
Time slots per episode & $T_{\max}$ & $100$ \\
\hline
\end{tabular}
\end{table}

We compare the proposed QR-D3QN method with the following baselines:
\begin{itemize}
    \item \emph{DQN}: Standard Deep Q Network that learns the expected return as a scalar value function.
    \item \emph{D3QN}: Builds on DQN with Double Q-learning and a dueling architecture to improve stability and action-value estimation, still based on expected returns.
    \item \emph{QR-DQN}: Distributional DRL algorithm using quantile regression to model the return distribution, without Double Q-learning or dueling structures.
    \item \emph{QR-D3QN}: The proposed approach combining quantile-based distributional learning with Double Q-learning and dueling architectures, designed to capture rare but persistent violations while maintaining stable learning.
    \item \emph{Drift-Plus-Penalty (DPP) Policy}: DPP is a well-established Lyapunov optimization framework for constrained stochastic optimization in networked systems and has been widely applied in AoI scheduling problems. Unlike DRL methods, DPP exploits system structure and does not rely on value function approximation, thereby providing a complementary benchmark from a classical control and operations research perspective. In this work, the resulting policy serves as a model-based heuristic benchmark derived from Lyapunov optimization principles. The details of the DPP policy are provided in Appendix \ref{sec:appendix_dpp}.
\end{itemize}

\begin{figure*}[t]
\centerline{\includegraphics[width=1\textwidth]{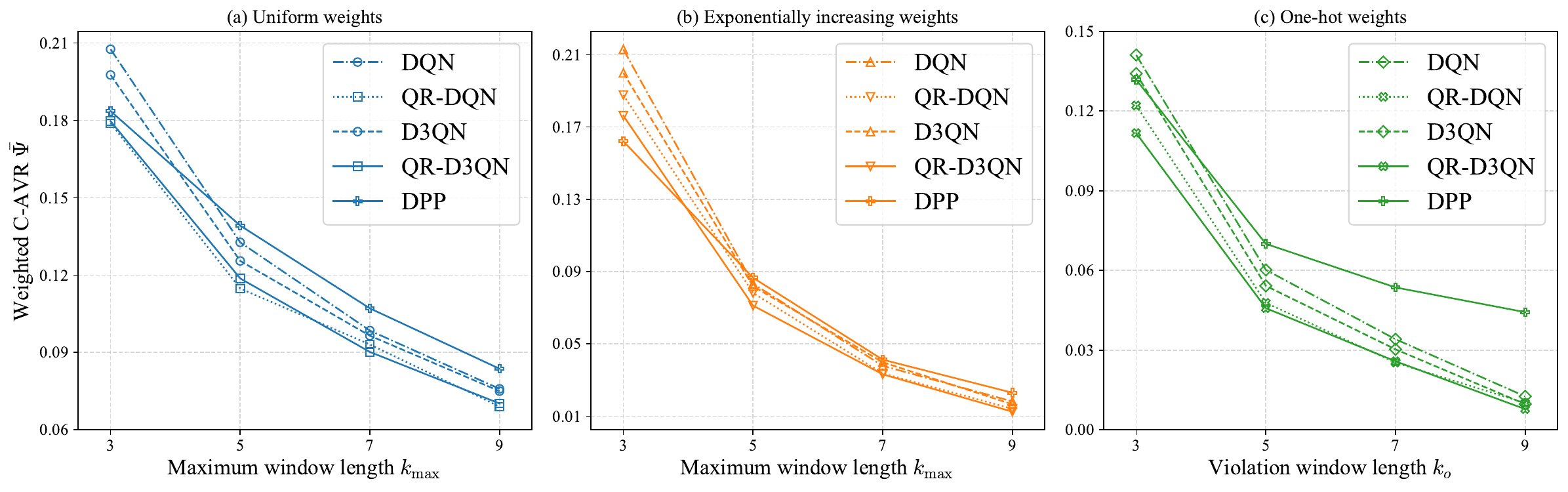}}
\caption{Weighted C-AVR performance $\bar{\Psi}$ of DQN, D3QN, QR-DQN, QR-D3QN, and DPP under (a) uniform weights, (b) exponential weights, and (c) one-hot weights. The horizontal axes represent the maximum window length $k_{\max}$ in (a) and (b), and the violation window length $k_o$ in (c) ($M=10$, $p_g^m=0.7$, $p_s^m=0.7$, and $\zeta=15$).}
\label{fig-weight_k_max_9_C-AVR}
\end{figure*}

\subsection{C-AVR Performance under Different Weighting Schemes}\label{Sec-PE2}
\subsubsection{Weighted C-AVR Performance} Fig.~\ref{fig-weight_k_max_9_C-AVR} shows the weighted C-AVR performance under different weighting schemes. Specifically, weighted C-AVR is plotted as a function of $k_{\max}$ under uniform and exponential weights in Fig. \ref{fig-weight_k_max_9_C-AVR}(a) and (b), respectively. For the one-hot setting, the objective reduces to a single-window C-AVR with window length $k_o$. To facilitate comparison, we set $k_o = k_{\max}$ and $w_{k_o}=1$, and plot the resulting performance as a function of $k_o$ in Fig.~\ref{fig-weight_k_max_9_C-AVR}(c). Under different weighting schemes, the weighted C-AVR decreases monotonically as $k_{\max}$ increases for all methods. This trend follows directly from the definition of C-AVR: longer violation windows occur less frequently and thus contribute smaller values to the aggregated metric. Considering the one-hot setting in Fig.~\ref{fig-weight_k_max_9_C-AVR}(c), for example, increasing $k_o$ gradually shifts the objective from a frequency-oriented metric ($k_o=1$, equivalent to AVR) to a persistence-sensitive reliability metric ($k_o>1$). The observed monotonic decrease with $k_o$ is therefore primarily induced by the metric itself rather than by differences in control policies.

Beyond this metric-induced effect, Fig.~\ref{fig-weight_k_max_9_C-AVR} shows a consistent performance ordering across different DRL algorithms. The two distributional methods (QR-DQN and QR-D3QN) achieve uniformly lower weighted C-AVR than the expectation-based baselines (DQN and D3QN). This indicates that weighted C-AVR optimization is inherently sensitive to the temporal persistence of AoI violations, and that learning the full return distribution provides a more informative learning signal than relying solely on expected returns. Moreover, the performance gap between QR-DQN and QR-D3QN is relatively modest, suggesting that the dominant improvement arises from distributional value representation, while Double Q-learning and the dueling architecture mainly contribute to stability and refinement of action evaluation. 

In addition, we include a Lyapunov-based DPP baseline in Fig.~\ref{fig-weight_k_max_9_C-AVR} to provide a structure-aware benchmark. When $k_{\max}$ (or $k_o$ in the one-hot case) is small, the weighted C-AVR objective is dominated by short-term violations and becomes closely aligned with instantaneous or near-term penalties. In this regime, by directly minimizing a one-step drift-plus-penalty surrogate, DPP captures the problem structure effectively, and DRL-based methods exhibit comparable performance. However, as $k_{\max}$ increases, the objective increasingly reflects long-horizon, persistence-driven penalties that depend on the evolution of violation sequences over multiple time slots. In this case, DPP's myopic structure becomes suboptimal, as it cannot explicitly account for long-term temporal dependencies. In contrast, DRL methods, particularly distributional ones, propagate delayed penalties through the value function and can therefore better anticipate and suppress persistent violation patterns. This explains why QR-D3QN increasingly outperforms DPP as $k_{\max}$ grows. All these results highlight that C-AVR emphasizes controlling persistent AoI violations rather than isolated threshold crossings. 

\begin{figure}[t]
\centerline{\includegraphics[width=0.49\textwidth]{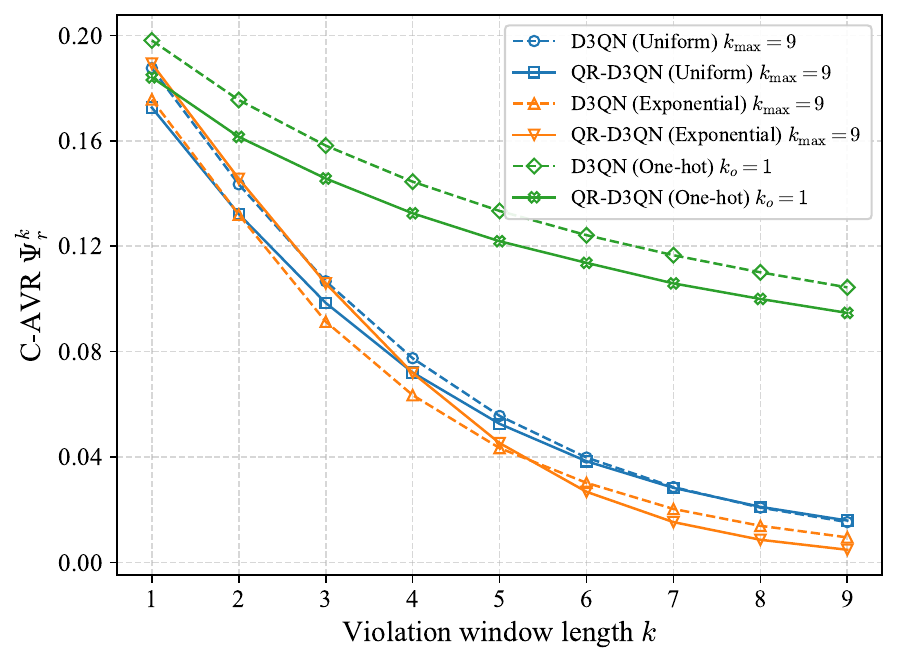}}
\caption{Per-component C-AVR $\Psi_r^k$ versus violation window length $k$ under different weighting schemes. Uniform and exponential weighting use $k_{\max}=9$, while the one-hot setting uses $k_o=1$ ($M=10$, $p_g^m=0.7$, $p_s^m=0.7$, and $\zeta=15$).}
\label{fig-k_1-9_C-AVR}
\end{figure}

\subsubsection{C-AVR Components Across Multiple Persistence Scales} While Fig. \ref{fig-weight_k_max_9_C-AVR} provides an aggregated evaluation, it does not reveal how performance is distributed across persistence scales. To this end, Fig. \ref{fig-k_1-9_C-AVR} examines the full C-AVR vector by reporting $\Psi_r^k$ for $k=\{1,\ldots,9\}$. Here, $k_{\max}=9$ is used for uniform and exponential weights. At the same time, $k_o=1$ is adopted in the one-hot case, corresponding to AVR optimization, a special case of the proposed weighted C-AVR framework. It is important to emphasize that Fig.~\ref{fig-k_1-9_C-AVR} evaluates the \emph{individual components} of the C-AVR vector, whereas the optimization objective is the aggregated metric $\bar{\Psi}$. Therefore, this figure serves as a diagnostic tool to assess how different objectives influence reliability across multiple persistence scales. Notably, even when training is performed under the one-hot setting (i.e., optimizing $\Psi_r^1$), we can still evaluate $\Psi_r^k$ for all $k$ by collecting the corresponding statistics under the learned policy.

As expected, $\Psi_r^k$ decreases monotonically with $k$ for all methods, reflecting the decreasing likelihood of long consecutive violations. More importantly, under uniform and one-hot weighting, QR-D3QN consistently achieves lower $\Psi_r^k$ than D3QN across almost all $k$. This demonstrates that the performance gain of distributional value learning is not restricted to a particular persistence scale but instead extends uniformly across the full spectrum of violation durations.

Under exponential weighting, an interesting trade-off is observed when examining the individual components of the C-AVR vector. Specifically, D3QN may achieve slightly lower $\Psi_r^k$ for small $k$ (e.g., $k<5$), whereas QR-D3QN significantly outperforms D3QN for larger $k$. This behavior is consistent with the structure of the objective, where exponential weights amplify the contribution of large-$k$ violations, leading the reward signal to be dominated by rare but high-impact events. Distributional methods retain quantile-level information about low-return trajectories and can therefore provide richer training signals for these rare penalties. In contrast, expectation-based methods may average out such events and instead allocate more resources to controlling short-term violations, which explains their relative advantage at small $k$. Consequently, as $k$ increases, the benefit of distributional value learning becomes more significant, and exponential weighting yields the lowest individual C-AVR $\Psi_r^k$ compared to other weighting schemes. Unlike exponential weighting, uniform weighting distributes emphasis evenly across all $k$, which is usually sufficient for short-to-moderate persistence scales.

Note in Fig.~\ref{fig-k_1-9_C-AVR} that the one-hot weighting scheme exhibits inferior performance for large $k$ compared to both uniform and exponential weighting. This indicates that a policy optimized solely for AVR ($k_o=1$) does not generalize to persistence-sensitive reliability metrics. More broadly, optimizing a single C-AVR component cannot guarantee good performance across other persistence scales. In contrast, the proposed weighted C-AVR objective provides a unified mechanism that not only optimizes an aggregate reliability measure but also improves individual components of the C-AVR vector in a coordinated manner.

\begin{figure}[t]
\centerline{\includegraphics[width=0.5\textwidth]{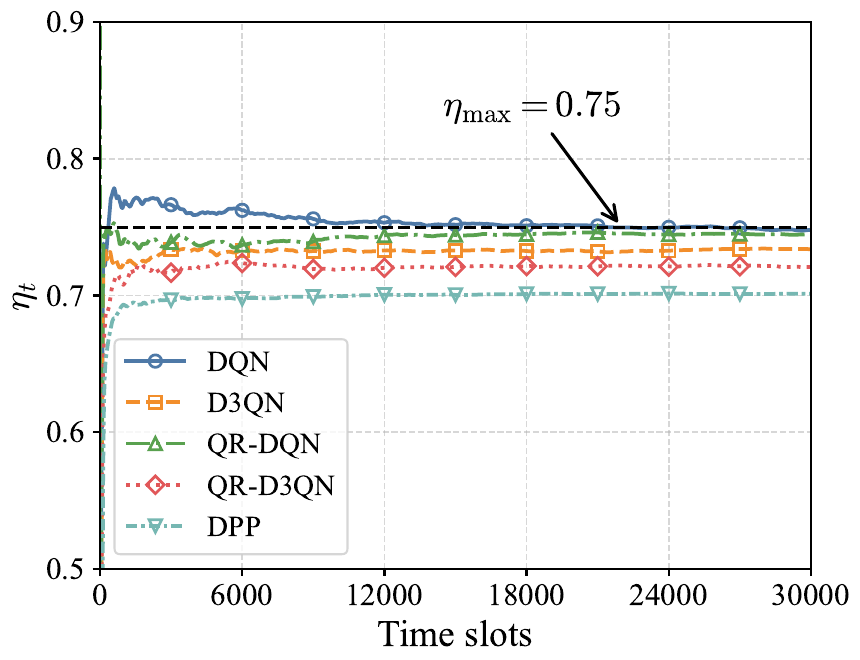}}
\caption{Average transmission cost over time under the cost constraint $\eta_{\max}=0.75$ with exponential weighting ($M=10$, $p_g^m=0.7$, $p_s^m=0.7$, and $\zeta=15$).}
\label{fig-Tcost}
\end{figure}

\subsubsection{Verification of Constraint Satisfaction} To verify constraint satisfaction, Fig.~\ref{fig-Tcost} depicts the evolution of the average transmission cost for all algorithms under the exponential weighting scheme as an example. All methods satisfy the prescribed constraint $\eta_{\max}=0.75$, indicating the feasibility of the learned policies. Importantly, the improved weighted C-AVR performance of QR-D3QN is achieved without increasing resource consumption. This confirms that distributional value learning enhances persistence-aware reliability through improved decision quality rather than more aggressive resource usage. In the following subsection, we further quantify these advantages by comparing distributional and expectation-based methods across varying system parameters.

\begin{figure*}[t]
\centerline{\includegraphics[width=1\textwidth]{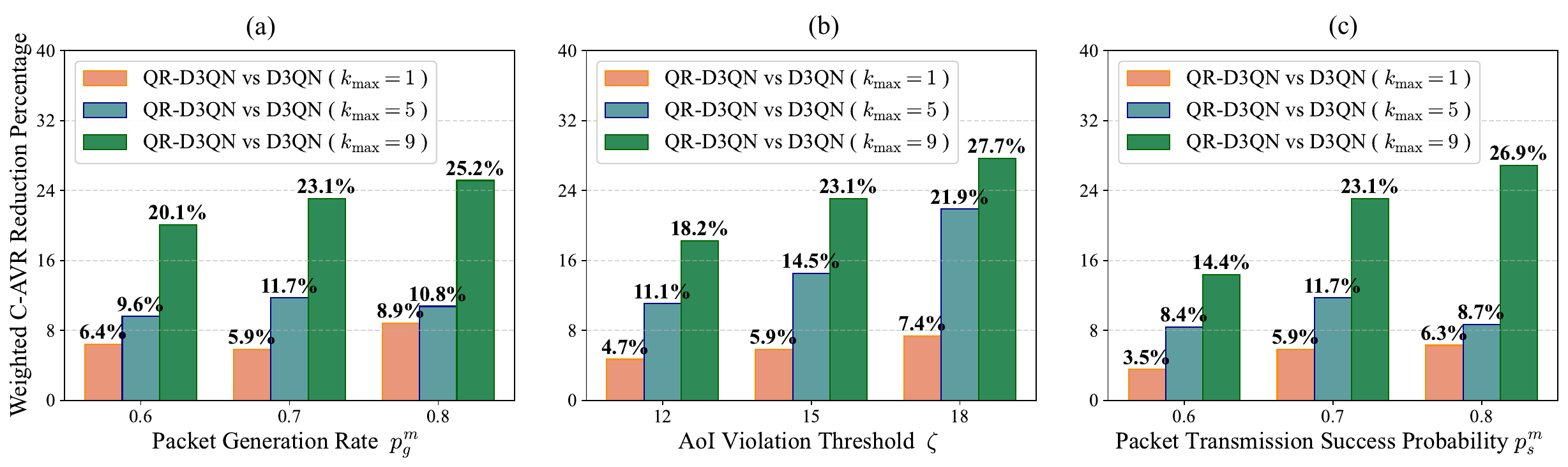}}
\caption{Relative weighted C-AVR reduction (\%) of QR-D3QN over D3QN for $k_{\max}\in\{1,5,9\}$ under exponential weighting and different (a) packet generation rates, (b) AoI violation thresholds, and (c) transmission successful probabilities.}
\label{fig-impro_159}
\end{figure*}

\subsection{Distributional DRL Performance under Different System Parameters}\label{Sec-P3}
Fig. \ref{fig-impro_159} quantifies the benefit of distributional value modeling under the weighted C-AVR objective. Specifically, it reports the relative reduction in weighted C-AVR achieved by QR-D3QN over D3QN for $k_{\max}\in\{1,5,9\}$ under exponential weighting, while varying key system parameters. Since QR-D3QN and D3QN share the same backbone network structure and training framework, while differing mainly in their value representations (distributional versus expectation-based), this comparison isolates the impact of modeling the full return distribution.  Across all subfigures, namely (a) packet generation rate, (b) AoI violation threshold, and (c) transmission success probability, QR-D3QN consistently outperforms D3QN. Notably, the performance gain increases with $k_{\max}$, indicating that distributional learning becomes increasingly advantageous as the objective places greater emphasis on persistence-driven violations.

In Fig. \ref{fig-impro_159}(a), as the packet generation rate increases, the performance gain of QR-D3QN becomes more significant, exceeding $20\%$ for $k_{\max}=9$ while remaining moderate for $k_{\max}=1$. When the generation rate is high, packet availability is no longer the limiting factor; instead, timely delivery under the transmission cost constraint becomes the dominant factor. In this regime, sequences of non-transmission or unsuccessful transmissions lead to uninterrupted AoI growth, thereby increasing the likelihood of consecutive violations. Consequently, the performance bottleneck shifts from average behavior to the control of such persistence events, amplifying the role of the tail in the weighted C-AVR objective and favoring distributional value learning. Next, Fig. \ref{fig-impro_159}(b) shows that as the AoI violation threshold $\zeta$ decreases, the relative improvement of QR-D3QN over D3QN diminishes. With stricter thresholds, violations occur more frequently across all policies, reducing the relative contribution of rare and persistent violation events in the weighted objective. Hence, the advantage of modeling the return distribution is partially attenuated. Nevertheless, QR-D3QN maintains consistent gains across all threshold settings, indicating that distributional value modeling remains beneficial even when violations are more common. Finally, Fig. \ref{fig-impro_159}(c) shows that as channel reliability improves, the relative gain of QR-D3QN increases, particularly for larger $k_{\max}$. A higher transmission success probability enhances the system's ability to control AoI evolution, enabling proactive prevention of long violation sequences. In this regime, performance increasingly depends on anticipating and avoiding rare persistence events. Distributional value learning offers an advantage by retaining quantile-level information about low-return trajectories during training, thereby enabling more effective suppression of such events.

\begin{figure*}[t]
\centerline{\includegraphics[width=0.9\textwidth]{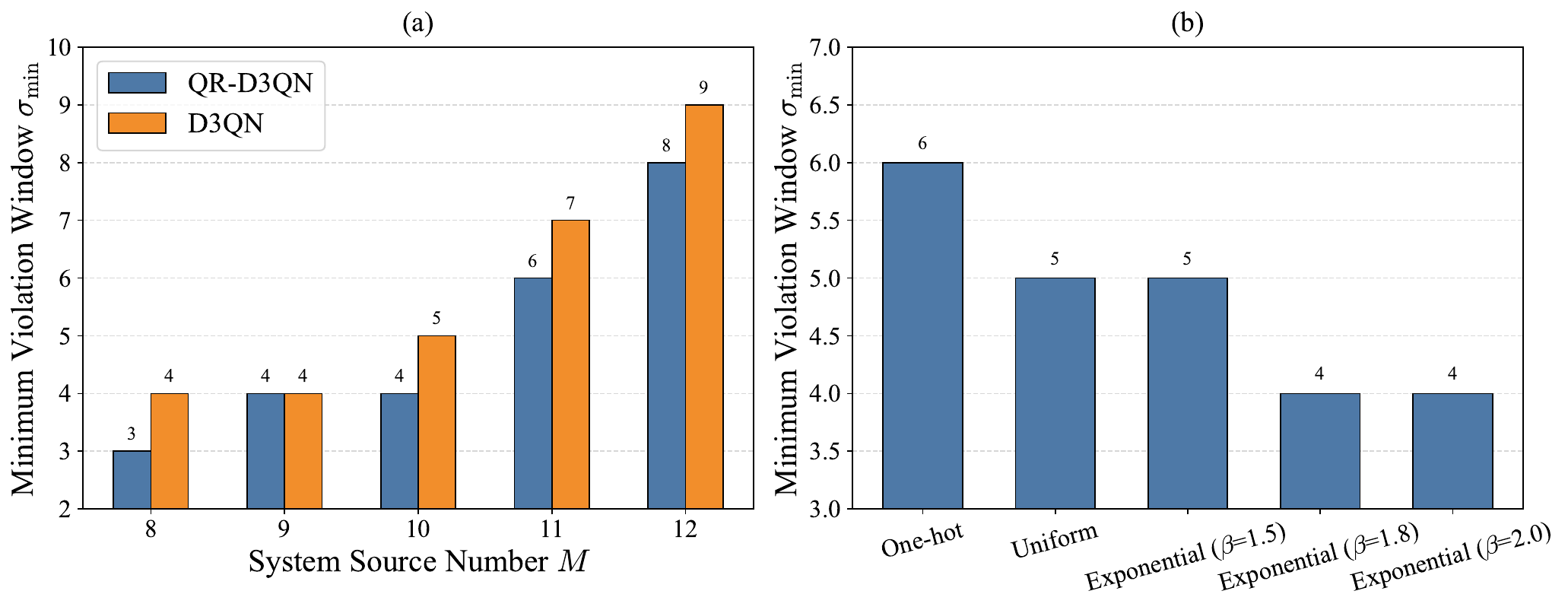}}
\caption{Tail persistence index $\sigma_{\min}$ under heterogeneous wireless environments: (a) comparison of $\sigma_{\min}$ between QR-D3QN and D3QN as the number of sources $M$ increases; (b) $\sigma_{\min}$ of QR-D3QN under different weighting schemes with $M=10$. For each source $m$, the packet generation probability $p_g^m$ and transmission success probability $p_s^m$ are independently sampled from $[0.6,0.8]$. $k_{\max}=9$ and $\hat{\epsilon}=0.05$; for the one-hot weighting scheme in (b), $k_o=9$.}
\label{fig-kmin}
\end{figure*}

\subsection{Tail Persistence Index: $\sigma_{\min}$ Analysis}\label{Sec-PE4}
While the weighted C-AVR and the detailed C-AVR vector characterize persistence-aware reliability from aggregate and profile perspectives, they do not directly provide a guarantee-oriented interpretation. From a system design viewpoint, it is often desirable to determine the minimum persistence scale beyond which the reliability requirement is consistently satisfied. To this end, we introduce the tail persistence index $\sigma_{\min}$, defined as
\begin{equation}
\sigma_{\min} = \min \left\{ k \in \{1,\dots,k_{\max}\} : \Psi_r^k \le \hat{\epsilon} \right\}.
\end{equation}
where $\hat{\epsilon}$ denotes a target violation level. If no persistence scale satisfies the target reliability level, we set $\sigma_{\min}=k_{\max}+1$, indicating that the violation probability remains above $\hat{\epsilon}$ throughout the considered persistence range. A smaller $\sigma_{\min}$ implies that the policy can suppress consecutive AoI violations below the target level starting from shorter persistence windows, thereby providing a stronger persistence-level reliability guarantee. Fig.~\ref{fig-kmin} evaluates $\sigma_{\min}$ under heterogeneous wireless environments. Specifically, for each source $m$, the packet generation probability $p_g^m$ and transmission success probability $p_s^m$ are independently sampled from the interval $[0.6,0.8]$. We set $k_{\max}=9$ and $\hat{\epsilon}=0.05$. For the one-hot weighting scheme in Fig.~\ref{fig-kmin}(b), we further set $k_o=9$.

Fig.~\ref{fig-kmin}(a) illustrates the impact of the number of sources $M$ on $\sigma_{\min}$. As the system size increases, QR-D3QN consistently achieves a smaller or equal $\sigma_{\min}$ compared with D3QN. Under heterogeneous traffic and channel conditions, different sources contribute unevenly to persistence-level violations, leading to a more skewed and high-variance return distribution. By explicitly modeling this distribution, QR-D3QN can better distinguish actions with similar expected returns but different long-term reliability implications, thereby achieving stronger persistence-aware guarantees. This result highlights that the proposed tail persistence index $\sigma_{\min}$ provides a direct and interpretable characterization of persistence-level reliability beyond aggregate weighted C-AVR metrics. 

Fig.~\ref{fig-kmin}(b) further examines the impact of the weighting scheme on $\sigma_{\min}$ under a fixed system size $M=10$. Since Fig.~\ref{fig-kmin}(a) already demonstrates the superiority of QR-D3QN over D3QN, only QR-D3QN is considered here. The results show that the weighting design significantly influences persistence-level reliability. In particular, the one-hot weighting scheme yields the largest $\sigma_{\min}$, indicating that optimizing a single persistence scale does not generalize well to broader persistence-aware reliability objectives. In contrast, both uniform and exponential weighting schemes achieve smaller $\sigma_{\min}$. In particular, exponential weighting places greater emphasis on long violation windows, thereby encouraging the learned policy to suppress persistent AoI violations more aggressively. This observation is consistent with the earlier C-AVR vector analysis, where exponential weighting achieved the most significant improvements for large persistence lengths. Overall, the proposed tail persistence index $\sigma_{\min}$ offers a complementary guarantee-oriented perspective by explicitly quantifying the minimum persistence scale at which a target reliability requirement is satisfied. The results further confirm that distributional reinforcement learning is particularly effective at improving tail-persistence guarantees in heterogeneous wireless status update systems.

\section{Conclusion}\label{Sec-Con}
We have investigated a reliability-aware scheduling for wireless status update systems, emphasizing the limitations of conventional metrics such as average AoI and AVR in capturing persistent information staleness. Specifically, we put forth the C-AVR vector, which characterizes AoI violation probabilities across multiple persistence scales, along with a weighted formulation that enables flexible prioritization of different violation durations, thereby providing a unified and fine-grained representation of timeliness reliability.

Optimizing the weighted C-AVR objective is challenging due to the dominance of rare but high-impact long violation sequences and the resulting skewed return distributions. To tackle this problem, we formulated the scheduling task as a CMDP and developed a distributional reinforcement learning approach based on QR-D3QN. By modeling a quantile-based return distribution rather than only a scalar expectation during training, the proposed method provides richer value-estimation signals for trajectories associated with persistent violations, while satisfying transmission cost constraints.

Extensive simulations validate both the proposed C-AVR vector framework and the distributional learning approach. The results show that distributional methods consistently outperform expectation-based baselines in optimizing weighted C-AVR, with gains becoming more significant as the objective emphasizes long violation windows. Moreover, a detailed per-component analysis further reveals that these improvements extend across persistence scales and are particularly significant for large $k$, confirming the effectiveness of distributional value modeling in suppressing prolonged violation sequences. The comparison across weighting schemes indicates that exponential weighting enhances tail suppression. Overall, these results establish the C-AVR vector as a principled framework for persistence-aware reliability evaluation and highlight the benefit of distributional value learning when system performance is strongly affected by rare but persistent events.

\appendices
\section{Drift-Plus-Penalty (DPP) Policy for Weighted C-AVR Scheduling}\label{sec:appendix_dpp}

This appendix presents the drift-plus-penalty (DPP) policy used in the simulations, which serves as a model-based heuristic benchmark derived from Lyapunov optimization principles. Unlike the proposed DRL approaches, the DPP policy exploits explicit system dynamics and performs online scheduling via a one-step optimization rule without value-function learning.

The weighted C-AVR objective characterizes the long-term persistence behavior of consecutive AoI violations across multiple temporal scales. To apply Lyapunov optimization, this objective must first be rewritten as an equivalent time-average penalty process. Recall that the violation counter $v(t,m)$ records the number of consecutive AoI violations of source $m$ at slot $t$. Define
\begin{equation}
\label{eq:dpp_H}
H(n)=\sum_{k=1}^{k_{\max}} w_k \mathbbm{1}[n\geq k].
\end{equation}
Using the equivalence $\mathbbm{1}[\mathcal{V}_t^k(m)]=\mathbbm{1}[v(t,m)\geq k]$, where $\mathcal{V}_t^k(m)$ denotes the $k$-slot consecutive violation event defined in the main paper, the instantaneous weighted persistence penalty can be written as
\begin{equation}
\label{eq:dpp_g}
g(t)
=
\frac{1}{M}
\sum_{m=1}^{M}
H(v(t,m)).
\end{equation}
This expression measures the aggregate persistence-aware reliability penalty incurred at slot $t$. Since $H(v(t,m))$ accumulates contributions from all persistence scales up to the current violation length, minimizing the long-term average of $g(t)$ suppresses both frequent short violations and prolonged violation runs.

Since $k_{\max}$ is finite and all terms are bounded, the finite summation over $k$, expectation, and time-average ($\limsup$) can be interchanged without affecting the asymptotic value. In addition, the difference between denominators $T$ and $T-k+1$ is a boundary term that vanishes as $T\to\infty$. Therefore, minimizing the long-term average of $g(t)$ is asymptotically equivalent to minimizing the weighted C-AVR objective, i.e.,
\begin{equation}
\bar{\Psi}
=
\limsup_{T\rightarrow\infty}
\frac{1}{T}
\sum_{t=1}^{T}
\mathbb{E}[g(t)].
\end{equation}
To enforce the average transmission cost constraint in \eqref{Problem}, we introduce the virtual queue
\begin{equation}
Z(t+1)
=
\max\left\{
Z(t)+c(t)-\eta_{\max},
0
\right\}.
\label{eq:appendix_virtual_queue}
\end{equation}
where $c(t)=\mathbbm{1}[a_t\ne 0]$ and $Z(1)=0$, following the time indexing $t=1,2,\ldots$ used in the main paper. If $Z(t)$ is mean-rate stable, then the average transmission-cost constraint is satisfied. Indeed, from \eqref{eq:appendix_virtual_queue},
\begin{align}
Z(t+1)\ge Z(t)+c(t)-\eta_{\max}.
\end{align}
Summing over $t=1,\ldots,T$ gives
\begin{align}
\frac{1}{T}\sum_{t=1}^{T}c(t)
\le \eta_{\max}+\frac{Z(T+1)-Z(1)}{T}.
\end{align}
Therefore, if $\lim_{T\to\infty}Z(T)/T=0$, then $\eta\le \eta_{\max}$. The virtual queue accumulates deviations from the desired average transmission budget. If the scheduler transmits excessively over time, $Z(t)$ increases, thereby discouraging future transmissions. Stabilizing the virtual queue therefore approximately enforces the long-term transmission cost constraint.

Let the Lyapunov function be
\begin{align}
L(Z(t))=\frac{1}{2}Z^2(t).
\label{eq:appendix_lyapunov}
\end{align}
The one-slot conditional Lyapunov drift is defined as
\begin{align}
\Delta(t)=\mathbb{E}\left[L(Z(t+1))-L(Z(t))\mid s_t,Z(t)\right].
\label{eq:appendix_drift_def}
\end{align}
From \eqref{eq:appendix_virtual_queue}, we have
\begin{align}
Z^2(t+1)
&\le \left(Z(t)+c(t)-\eta_{\max}\right)^2 \\
&=Z^2(t)+\left(c(t)-\eta_{\max}\right)^2
+2Z(t)\left(c(t)-\eta_{\max}\right).
\label{eq:appendix_square_bound}
\end{align}
Since $c(t)\in\{0,1\}$, there exists a finite constant $B_Z$ such that
\begin{align}
\frac{1}{2}\left(c(t)-\eta_{\max}\right)^2\le B_Z.
\label{eq:appendix_BZ}
\end{align}
Therefore,
\begin{align}
\Delta(t)
\le B_Z+Z(t)\mathbb{E}\left[c(t)-\eta_{\max}\mid s_t,Z(t)\right].
\label{eq:appendix_drift_bound}
\end{align}

Following the Lyapunov optimization framework, DPP minimizes an upper bound on the one-slot drift-plus-penalty function
\begin{align}
\Delta(t)+V_{\mathrm{DPP}}\mathbb{E}\left[g(t+1)\mid s_t,Z(t)\right],
\label{eq:appendix_dpp_objective}
\end{align}
where $V_{\mathrm{DPP}}>0$ controls the tradeoff between the weighted C-AVR penalty and the virtual-queue pressure. Substituting \eqref{eq:appendix_drift_bound} into \eqref{eq:appendix_dpp_objective} and removing terms independent of the action, the per-slot DPP decision is
\begin{align}
a_t^*=\arg\min_{a\in\{0,1,\ldots,M\}}
\left\{
Z(t)\mathbbm{1}[{a\neq 0}]
+V_{\mathrm{DPP}}\mathbb{E}\left[g(t+1)\mid s_t,a\right]
\right\}.
\label{eq:appendix_dpp_min}
\end{align}
Unlike exact dynamic programming approaches, the DPP method only minimizes a one-step conditional surrogate rather than the full long-term objective. The resulting policy should therefore be interpreted as a heuristic online control benchmark derived from Lyapunov optimization principles. 

We next derive the corresponding scheduling index. Suppose source $m$ is considered for transmission at slot $t$. If no successful update occurs, the next-slot receiver-side AoI becomes
\begin{equation}
d_{m,F}^{r}
=
\min\{
\Delta_r(t,m)+1,
\Delta_{\max}
\}.
\end{equation}
If the transmission succeeds, the next-slot receiver-side AoI becomes
\begin{equation}
d_{m,S}^{r}
=
\min\{
\Delta_s(t,m)+1,
\Delta_{\max}
\},
\end{equation}
where $\Delta_r(t,m)$ and $\Delta_s(t,m)$ denote the receiver-side and transmitter-side AoI, respectively. The corresponding next-slot violation counters are
\begin{equation}
\label{eq:dpp_counter_F}
F_m=
\begin{cases}
\min\{v(t,m)+1,k_{\max}\}, & d_{m,F}^{r}>\zeta,\\
0, & d_{m,F}^{r}\leq \zeta,
\end{cases}
\end{equation}
and
\begin{equation}
\label{eq:dpp_counter_S}
S_m=
\begin{cases}
\min\{v(t,m)+1,k_{\max}\}, & d_{m,S}^{r}>\zeta,\\
0, & d_{m,S}^{r}\leq \zeta.
\end{cases}
\end{equation}
where $\zeta$ is the predefined AoI violation threshold. Equation \eqref{eq:dpp_counter_F} corresponds to the case where the violation process continues without successful refresh. In contrast, \eqref{eq:dpp_counter_S} captures the post-update evolution after successful transmission. Importantly, a successful transmission does not necessarily terminate the violation process because the delivered packet itself may already be stale, i.e., $\Delta_s(t,m)+1>\zeta$. In such cases, the consecutive violation counter continues to increase even after successful delivery.

A successful update therefore reduces the next-slot weighted persistence penalty from $H(F_m)$ to $H(S_m)$. Taking the transmission success probability $p_s^m$ into account, the expected persistence reduction achieved by scheduling source $m$ is
\begin{equation}
\label{eq:dpp_index}
I_m(t)
=
p_s^m
\left[
H(F_m)-H(S_m)
\right].
\end{equation}
The quantity $I_m(t)$ can therefore be interpreted as a persistence-aware scheduling index that measures the expected reduction in future weighted violation severity achieved by transmitting source $m$. The source providing the largest expected reduction in weighted persistence penalty is selected as
\begin{equation}
\label{eq:dpp_best}
m^*
=
\arg\max_{m\in \{1,2,...,M\}}
I_m(t).
\end{equation}

Under the drift-plus-penalty rule in \eqref{eq:appendix_dpp_min}, scheduling source $m^*$ is preferred to remaining idle whenever
\begin{equation}
\frac{V_{\mathrm{DPP}}}{M}
I_{m^*}(t)
>
Z(t).
\end{equation}
This condition admits an intuitive interpretation: transmission is performed only when the expected persistence reduction outweighs the current virtual congestion level induced by the transmission budget constraint. Since the factor $1/M$ is a positive constant independent of the action, it can be absorbed into the control parameter $V_{\mathrm{DPP}}$. The resulting DPP scheduling rule becomes
\begin{equation}
\label{eq:dpp_policy}
a_t=
\begin{cases}
m^*, &
V_{\mathrm{DPP}} I_{m^*}>Z(t),
\\
0, &
\text{otherwise}.
\end{cases}
\end{equation}
The resulting DPP policy is summarized in Algorithm~\ref{alg:dpp}. In our simulations, we set $V_{\mathrm{DPP}}=10$.

\begin{algorithm}[t]
\caption{DPP Policy for Weighted C-AVR Scheduling}
\label{alg:dpp}
\begin{algorithmic}[1]
\REQUIRE Weight vector $\mathbf{W}=(w_1,\ldots,w_{k_{\max}})$, threshold $\zeta$, AoI bound $\Delta_{\max}$, cost budget $\eta_{\max}$, success probabilities $\{p_s^m\}_{m\in\{1,2,...,M\}}$, and DPP parameter $V_{\mathrm{DPP}}$ after absorbing the constant factor $1/M$.
\STATE Initialize $Z(1)=0$, $\Delta_s(0,m)=0$, $\Delta_r(1,m)=1$, and $v(1,m)=0$ for all $m\in\{1,2,...,M\}$
\STATE Construct $H(n)=\sum_{k=1}^{k_{\max}} w_k\mathbbm{1}\left[{n\ge k}\right]$ for $n=0,1,\ldots,k_{\max}$
\FOR{each time slot $t=1,2,\ldots$}
    \STATE Update $\Delta_s(t,m)$ according to packet arrivals for all $m\in\{1,2,...,M\}$
    \STATE Observe $s_t$ and the virtual queue $Z(t)$
    \FOR{each source $m\in\{1,2,...,M\}$}
        \STATE Compute $F_m$ using \eqref{eq:dpp_counter_F}
        \STATE Compute $S_m$ using \eqref{eq:dpp_counter_S}
        \STATE Compute $I_m(t)=p_s^m (H(F_m)-H(S_m))$
    \ENDFOR
    \STATE Select $m^*=\arg\max_{m\in\{1,2,...,M\}} I_m(t)$
    \IF{$V_{\mathrm{DPP}}I_{m^*}(t)>Z(t)$}
        \STATE Schedule source $m^*$, i.e., set $a_t=m^*$
    \ELSE
        \STATE Stay idle, i.e., set $a_t=0$
    \ENDIF
    \STATE Execute $a_t$ and observe the transmission outcome if a source is scheduled
    \STATE Update $\Delta_r(t+1,m)$ and $v(t+1,m)$ for all $m \in \{1,2,...,M\}$
    \STATE Update $Z(t+1)$ by \eqref{eq:appendix_virtual_queue}
\ENDFOR
\end{algorithmic}
\end{algorithm}

\bibliographystyle{IEEEtran}
\bibliography{refer}

\end{document}